# Derivatives of spin dynamics simulations

Ilya Kuprov†, Christopher T. Rodgers‡

†*Oxford e-Research Centre, University of Oxford,
7 Keble Road, Oxford, OX1 3QG, UK.*

‡*Oxford Centre for Clinical Magnetic Resonance Research,
Level 0, John Radcliffe Hospital, Oxford, OX3 9DU, UK.*

Email: ilyaATkuprovDOTcom , pubs-cATrodgersDOTorgDOTuk

**Abstract**

We report analytical equations for the derivatives of spin dynamics simulations with respect to pulse sequence and spin system parameters. The methods described are significantly faster, more accurate and more reliable than the finite difference approximations typically employed. The resulting derivatives may be used in fitting, optimization, performance evaluation and stability analysis of spin dynamics simulations and experiments.

# Introduction

The response of a magnetic resonance experiment to a small perturbation in the spin system or pulse sequence parameters is useful in two contexts: spectral fitting [1-5] and pulse sequence design using optimal control theories [6-9].

Spectral fitting algorithms require gradients for optimization step control [10-12]. In small-scale calculations, gradient-free methods [13-14] are adequate, but as the system gets bigger, parameters proliferate and such gradient-free methods become slow and hence impractical. As our colleagues in electronic structure theory quickly discovered [15-16], for large systems analytical gradients are required for efficient optimization.

In pulse sequence design using optimal control, an important question is the robustness of the resulting waveforms with respect to non-idealities in the control operators and external noise [17-18] as well as the natural variability in the practically encountered spin systems. A derivative of the coherence transfer efficiency with respect to *e.g.* a systematic phase distortion in the RF channel can be interpreted as a measure of stability of the pulse sequence with respect to that notorious instrumental problem.

Importantly, the derivatives in question often cannot be obtained numerically: modern large-scale simulation algorithms have multiple dynamic cut-offs and tolerances (in orthogonalization, matrix inversion, singular value decomposition and other essential mathematical procedures) meaning that a small perturbation in a parameter may trigger a step change in the simulation result, *e.g.* inclusion or rejection of a particular vector in the basis. In other words, many algorithms are not numerically differentiable with respect to their parameters. They may also display high levels of numerical noise due to the finite precision of machine arithmetic. Even when they are reasonably accurate, numerical derivatives have a high computational cost in such large-scale simulations: typically between two and four separate simulations per parameter.

We report in this communication the analytical equations for the derivatives of spin dynamics simulations with respect to both pulse sequence and spin system parameters. The equations reported are significantly faster, much more accurate, and, importantly, much more reliable than finite difference approximations.

## Theory

Spin dynamics simulations may be formulated either in Hilbert space, where operators are represented as $n \times n$ matrices, or in Liouville space, where operators are represented as $n^2$-element vectors and superoperators are $n^2 \times n^2$ matrices [19]. We describe the spin system in terms of its density operator $\hat{\rho}(t)$. In Hilbert space, this evolves according to the Liouville–von Neumann equation [19]:

$$\frac{\partial \hat{\rho}(t)}{\partial t} = -\mathrm{i}[\hat{H}(t), \hat{\rho}(t)], \tag{1}$$

in which $\hat{H}(t)$ is the Hamiltonian operator. In Liouville space, the density operator evolves under the system Liouvillian superoperator $\hat{\hat{L}}(t)$, which may include contributions from relaxation and chemical kinetics:

$$\frac{\partial \hat{\rho}(t)}{\partial t} = -\mathrm{i}\hat{\hat{L}}(t)\hat{\rho}(t). \tag{2}$$

Both $\hat{H}(t)$ and $\hat{\hat{L}}(t)$ are assumed to be piecewise continuous. Differentiation may be performed by several approaches in both representations.

### I. Time domain derivative superoperator

Starting in Liouville space, it is natural to seek a superoperator $\hat{\hat{A}}(t)$ that acts on the density operator giving its derivative $\hat{\rho}'_\alpha(t)$ with respect to a simulation parameter $\alpha$:

$$\hat{\hat{A}}(t) = \frac{\partial}{\partial \alpha} \quad \text{i.e.} \quad \hat{\rho}'_\alpha(t) = \hat{\hat{A}}(t)\hat{\rho}(t). \tag{3}$$

Integrating Equation (2) analytically for an infinitesimal time step $\Delta t$ yields

$$\hat{\rho}(t + \Delta t) = \exp\left[-\mathrm{i}\hat{\hat{L}}\Delta t\right]\hat{\rho}(t), \tag{4}$$

which is free from time derivatives (which do not necessarily commute with $\partial/\partial\alpha$) and may be directly differentiated with respect to the parameter:

$$\hat{\rho}'_\alpha(t+\Delta t) = \left(\frac{\partial}{\partial\alpha}\exp\left[-i\hat{\hat{L}}\Delta t\right]\right)\hat{\rho}(t) + \exp\left[-i\hat{\hat{L}}\Delta t\right]\hat{\rho}'_\alpha(t) \tag{5}$$

Combining Equation (5) with the definition given in Equation (3) yields a time propagation law for the differentiation superoperator:

$$\hat{\hat{A}}(t+\Delta t) = \left(\frac{\partial}{\partial\alpha}\exp\left[-i\hat{\hat{L}}\Delta t\right]\right)\exp\left[i\hat{\hat{L}}\Delta t\right] + \exp\left[-i\hat{\hat{L}}\Delta t\right]\hat{\hat{A}}(t)\exp\left[i\hat{\hat{L}}\Delta t\right] \tag{6}$$

Taking the limit $\Delta t \to 0$ and neglecting $O(\Delta t^2)$ and higher terms, gives the equation of motion for the differentiation superoperator:

$$\frac{\partial\hat{\hat{A}}}{\partial t} = -i[\hat{\hat{L}},\hat{\hat{A}}] - i\hat{\hat{L}}'_\alpha \tag{7}$$

where $\hat{\hat{L}}'_\alpha$ is a Liouvillian derivative with respect to the simulation parameter in question. The initial conditions are in most cases $\hat{\hat{A}}(0) = 0$ and $\hat{\rho}'_\alpha(0) = 0$, reflecting the fact that the simulation starts from a user-specified state that does not depend on $\alpha$. Even though the evolution of the differentiation superoperator is governed by the same Liouvillian, it is independent from the evolution of the density matrix.

Analogous reasoning yields the following equations for the second derivative superoperator $\hat{\hat{W}}$:

$$\hat{\hat{W}}(t+\Delta t) = \exp\left[-i\hat{\hat{L}}\Delta t\right]\hat{\hat{W}}(t)\exp\left[i\hat{\hat{L}}\Delta t\right] + \left(\frac{\partial}{\partial\alpha}\exp\left[-i\hat{\hat{L}}\Delta t\right]\right)\hat{\hat{B}}(t)\exp\left[i\hat{\hat{L}}\Delta t\right] + \\ + \left(\frac{\partial}{\partial\beta}\exp\left[-i\hat{\hat{L}}\Delta t\right]\right)\hat{\hat{A}}(t)\exp\left[i\hat{\hat{L}}\Delta t\right] + \left(\frac{\partial^2}{\partial\alpha\partial\beta}\exp\left[-i\hat{\hat{L}}\Delta t\right]\right)\exp\left[i\hat{\hat{L}}\Delta t\right] \tag{8}$$

$$\frac{\partial\hat{\hat{W}}}{\partial t} = -i[\hat{\hat{L}},\hat{\hat{W}}] - i\hat{\hat{L}}'_\alpha\hat{\hat{B}}(t) - i\hat{\hat{L}}'_\beta\hat{\hat{A}}(t) - i\hat{\hat{L}}''_{\alpha\beta} \tag{9}$$

where $\hat{\hat{B}}(t)$ is a first derivative superoperator with respect to the second parameter $\beta$. Higher order derivatives may be obtained in a similar way.

## II. Liouville-space derivative density matrix

A second approach identifies an equation of motion for the derivative of the density matrix in Liouville space. We begin by taking the $\Delta t \to 0$ limit of Equation (5), which yields the following equation of motion for the derivative density matrix

$$\frac{\partial \hat{\hat{\rho}}'_\alpha}{\partial t} = -i\hat{\hat{L}}\hat{\rho}'_\alpha - i\hat{\hat{L}}'_\alpha \hat{\rho} \tag{10}$$

This has the same general form as Equation (2) with the addition of a second driving term that depends on the density matrix itself.

Repeating these steps for the second derivative yields the following equations for the second derivative density matrix dynamics:

$$\hat{\rho}''_{\alpha\beta}(t+\Delta t) = \hat{\hat{P}}''_{\alpha\beta}\hat{\rho}(t) + \hat{\hat{P}}'_\alpha \hat{\rho}'_\beta(t) + \hat{\hat{P}}'_\beta \hat{\rho}'_\alpha(t) + \hat{\hat{P}}\hat{\rho}''_{\alpha\beta}(t); \quad \hat{\hat{P}} = \exp\left[-i\hat{\hat{L}}\Delta t\right] \tag{11}$$

$$\frac{\partial \hat{\rho}''_{\alpha\beta}}{\partial t} = -i\hat{\hat{L}}''_{\alpha\beta}\hat{\rho} - i\hat{\hat{L}}'_\alpha \hat{\rho}'_\beta - i\hat{\hat{L}}'_\beta \hat{\rho}'_\alpha - i\hat{\hat{L}}\hat{\rho}''_{\alpha\beta} \tag{12}$$

Higher order derivatives may be obtained in a similar manner.

## III. Hilbert-space derivative density matrix

It is also possible to find an equivalent equation of motion for the derivative of the density matrix in Hilbert space. Integrating Equation (1) for an infinitesimal time step $\Delta t$

$$\hat{\rho}(t+\Delta t) = \exp\left(-i\hat{H}\Delta t\right)\hat{\rho}(t)\exp\left(i\hat{H}\Delta t\right) \tag{13}$$

and differentiating with respect to the parameter $\alpha$ gives

$$\begin{aligned}\hat{\rho}'_\alpha(t+\Delta t) &= \left(\frac{\partial}{\partial \alpha}\exp\left[-i\hat{H}\Delta t\right]\right)\hat{\rho}(t)\exp\left[i\hat{H}\Delta t\right] + \\ &+ \exp\left[-i\hat{H}\Delta t\right]\hat{\rho}'_\alpha(t)\exp\left[i\hat{H}\Delta t\right] + \\ &+ \exp\left[-i\hat{H}\Delta t\right]\hat{\rho}(t)\left(\frac{\partial}{\partial \alpha}\exp\left[i\hat{H}\Delta t\right]\right).\end{aligned} \tag{14}$$

Taking the $\Delta t \to 0$ limit then yields the following Hilbert-space equation of motion for the derivative density matrix

$$\frac{\partial \hat{\rho}'_\alpha}{\partial t} = -i[\hat{H}, \hat{\rho}'_\alpha] - i[\hat{H}'_\alpha, \hat{\rho}] \qquad (15)$$

Like Equation (10) above, this expression for the derivative density matrix is the same as Equation (1) with an additional driving term that depends on the density matrix itself. Once again, higher derivatives may be obtained by analogous reasoning. For example, the second derivative density matrix satisfies

$$\hat{\rho}''_{\alpha\beta}(t+\Delta t) = \hat{\Lambda}''_{\alpha\beta}\hat{\rho}(t)\hat{\Pi} + \hat{\Lambda}'_\alpha \hat{\rho}'_\beta(t)\hat{\Pi} + \hat{\Lambda}'_\alpha \hat{\rho}(t)\hat{\Pi}'_\beta + \hat{\Lambda}'_\beta \hat{\rho}'_\alpha(t)\hat{\Pi} +$$
$$+\hat{\Lambda}\hat{\rho}''_{\alpha\beta}(t)\hat{\Pi} + \hat{\Lambda}\hat{\rho}'_\alpha(t)\hat{\Pi}'_\beta + \hat{\Lambda}'_\beta \hat{\rho}(t)\hat{\Pi}'_\alpha + \hat{\Lambda}\hat{\rho}'_\beta(t)\hat{\Pi}'_\alpha + \hat{\Lambda}\hat{\rho}(t)\hat{\Pi}''_{\alpha\beta}; \qquad (16)$$
$$\hat{\Lambda} = \exp\left[-i\hat{H}\Delta t\right]; \quad \hat{\Pi} = \exp\left[i\hat{H}\Delta t\right]$$

$$\frac{\partial \hat{\rho}''_{\alpha\beta}}{\partial t} = -i[\hat{H}''_{\alpha\beta}, \hat{\rho}] - i[\hat{H}'_\alpha, \hat{\rho}'_\beta] - i[\hat{H}'_\beta, \hat{\rho}'_\alpha] - i[\hat{H}, \hat{\rho}''_{\alpha\beta}] \qquad (17)$$

**IV. Evaluation of key derivatives**

The individual derivatives involved in the equations above are all very straightforward. The Hamiltonian and Liouvillian derivative with respect to a parameter returns an operator or superoperator corresponding to that parameter. For example, in Hilbert space the derivative of the Hamiltonian with respect to $\alpha = a_{nk}$ is

$$\hat{H} = \sum_i \omega_i \hat{S}_Z^{(i)} + \sum_{ij} a_{ij} \hat{\vec{S}}^{(i)} \cdot \hat{\vec{S}}^{(j)} + \ldots \quad \Rightarrow \quad \frac{\partial \hat{H}}{\partial a_{nk}} = \hat{\vec{S}}^{(n)} \cdot \hat{\vec{S}}^{(k)} \qquad (18)$$

And the derivative of the corresponding Liouvillian is

$$\hat{\hat{L}} = \hat{H} \otimes \hat{I} - \hat{I} \otimes \hat{H}^{\mathrm{T}}$$
$$\frac{\partial \hat{\hat{L}}}{\partial a_{nk}} = \frac{\partial \hat{H}}{\partial a_{nk}} \otimes \hat{I} - \hat{I} \otimes \left(\frac{\partial \hat{H}}{\partial a_{nk}}\right)^{\mathrm{T}} = \left(\hat{\vec{S}}^{(n)} \cdot \hat{\vec{S}}^{(k)}\right) \otimes \hat{I} - \hat{I} \otimes \left(\hat{\vec{S}}^{(n)} \cdot \hat{\vec{S}}^{(k)}\right)^{\mathrm{T}} \qquad (19)$$

The parameter derivative of the matrix exponential may be computed from its power series definition (or using more sophisticated techniques [20-22]), taking into account the fact that $\hat{\hat{L}}'_\alpha$ might not commute with $\hat{\hat{L}}$ and using sufficiently small time increments so that $\hat{\hat{L}}$ may be treated as being piecewise constant

$$\frac{\partial}{\partial \alpha} \exp\left(-i\hat{\hat{L}}\Delta t\right) = \sum_{n=1}^{\infty} \frac{(-i\Delta t)^n}{n!} \sum_{k=0}^{n-1} \hat{\hat{L}}^k \hat{\hat{L}}'_\alpha \hat{\hat{L}}^{n-k-1} \qquad (20)$$

and similarly for the Hamiltonian exponential. For very large sparse matrices encountered in spin dynamics this Taylor series, ironically, works best because matrix sparsity is usually inherited after multiplication, particularly if a clean-up pass ($A_{ik} \to 0$ for all $|A_{ik}| < \varepsilon$) is performed at each stage.

Scaling and squaring may be used if necessary to accelerate convergence. It is worth noting that the scaling and squaring procedure is different for the derivative exponential [20]:

$$\frac{\partial}{\partial \alpha}\exp\left(-i\hat{\hat{L}}\Delta t\right) = \exp\left(\frac{-i\hat{\hat{L}}\Delta t}{2}\right)\left(\frac{\partial}{\partial \alpha}\exp\left[\frac{-i\hat{\hat{L}}\Delta t}{2}\right]\right) + \left(\frac{\partial}{\partial \alpha}\exp\left[\frac{-i\hat{\hat{L}}\Delta t}{2}\right]\right)\exp\left(\frac{-i\hat{\hat{L}}\Delta t}{2}\right) \qquad (21)$$

as opposed to the standard exponential procedure:

$$\exp\left(-i\hat{\hat{L}}\Delta t\right) = \exp\left(\frac{-i\hat{\hat{L}}\Delta t}{2}\right)^2 \qquad (22)$$

Although the superoperator formulation given by Equation (7) and the equation of motion formulation given by Equations (10) and (15) are formally equivalent, the latter are likely to be more computationally efficient because Equations (5) and (14) do not require Liouville-space matrix-matrix multiplications, which do occur in Equation (6). It is important to note that all three time-domain formalisms outlined above are numerically stable and compatible with time-dependent Hamiltonians and Liouvillians.

## V. Frequency domain spectrum derivatives

In Liouville space simulations with a static Liouvillian superoperator, the positive-time Fourier transform of the Liouville–von Neumann equation is

$$\frac{\partial \hat{\rho}(t)}{\partial t} = -i\hat{\hat{L}}\hat{\rho}(t) \quad \Rightarrow \quad i\left(\omega\hat{\hat{I}} + \hat{\hat{L}}\right)\hat{\rho}(\omega) = \hat{\rho}_0 \tag{23}$$
$$\hat{\rho}(0) = \hat{\rho}_0$$

where $\hat{\rho}_0$ is the initial density matrix, $\omega$ is frequency, and $\hat{\hat{I}}$ is a unit matrix of the same dimensions as the Liouvillian superoperator. Differentiating with respect to the parameter $\alpha$ and rearranging yields

$$\hat{\rho}'_\alpha(\omega) = -\left(\omega\hat{\hat{I}} + \hat{\hat{L}}\right)^{-1} \hat{\hat{L}}'_\alpha \rho(\omega) \tag{24}$$

In most experiments and simulations, a specific observable – we shall call it $Q$ – is monitored. Hence, the quantity of interest is the derivative of $Q$:

$$Q'_\alpha(\omega) = \hat{Q}^\dagger \hat{\rho}'_\alpha(\omega) = \left\{-\hat{Q}^\dagger \left(\omega\hat{\hat{I}} + \hat{\hat{L}}\right)^{-1} \hat{\hat{L}}'_\alpha\right\} \hat{\rho}(\omega) = \hat{Q}'_\alpha(\omega)^\dagger \hat{\rho}(\omega) \tag{25}$$

$$\hat{Q}'_\alpha(\omega)^\dagger = -\hat{Q}^\dagger \left(\omega\hat{\hat{I}} + \hat{\hat{L}}\right)^{-1} \hat{\hat{L}}'_\alpha \tag{26}$$

where the dagger sign denotes a Hermitian conjugate. Once the "derivative observable row vector" $\hat{Q}'_\alpha(\omega)^\dagger$ has been computed, the derivative spectrum may be obtained quite simply with a vector–vector product followed by an inverse Fourier transform. Note that the row vector $\hat{Q}^\dagger(\omega\hat{\hat{I}} + \hat{\hat{L}})^{-1}$ may be computed directly [23] without ever requiring the time-consuming evaluation of the full $(\omega\hat{\hat{I}} + \hat{\hat{L}})^{-1}$ inverse. Large savings are also possible if derivatives with respect to other parameters are sought because the $\hat{Q}^\dagger(\omega\hat{\hat{I}} + \hat{\hat{L}})^{-1}$ intermediate row vector does not need to be recalculated. Also, specific points in the spectrum may be monitored without the need to simulate the time trajectory in its entirety.

Similar to the first derivative treatment above, the frequency domain equations for the second derivative density matrix Fourier transform are:

$$\begin{aligned}\hat{\rho}''_{\alpha\beta}(\omega) &= -\left(\omega\hat{\hat{I}} + \hat{\hat{L}}\right)^{-1}\left(\hat{\hat{L}}'_\alpha \hat{\rho}'_\beta(\omega) + \hat{\hat{L}}'_\beta \hat{\rho}'_\alpha(\omega) + \hat{\hat{L}}''_{\alpha\beta}\hat{\rho}(\omega)\right) \\ &= \left(\omega\hat{\hat{I}} + \hat{\hat{L}}\right)^{-1}\left(\hat{\hat{L}}'_\alpha\left(\omega + \hat{\hat{L}}\right)^{-1}\hat{\hat{L}}'_\beta + \hat{\hat{L}}'_\beta\left(\omega + \hat{\hat{L}}\right)^{-1}\hat{\hat{L}}'_\alpha - \hat{\hat{L}}''_{\alpha\beta}\right)\hat{\rho}(\omega)\end{aligned} \tag{27}$$

where the intermediate results could also be re-used to calculate further derivatives with respect to different parameters. It is worth noting that the extremely sparse structure of Liouvillian derivatives, which contain at most only a handful of non-zeros, means that the Hessian $\hat{\hat{L}}''_{\alpha\beta}$ in Equation (27) is very sparse.

## VI. Eigensystem differentiation

A complementary strategy for calculating derivatives with respect to a simulation parameter is available in systems that have a piecewise constant or time-independent Hamiltonian $\hat{H}$ that can be diagonalized reasonably quickly. We seek the derivative of an observable

$$\frac{\partial}{\partial \alpha}\langle \hat{Q}(t)\rangle = \frac{\partial}{\partial \alpha}\mathrm{Tr}\left[ e^{-i\hat{H}t}\hat{\rho}_0 e^{i\hat{H}t}\hat{Q}\right]. \tag{28}$$

in which the parameter derivatives of the initial density matrix $\hat{\rho}_0$ and the observable operator $\hat{Q}$ are typically zero or simple to determine. Equation (28) is easily evaluated using the product rule once the derivative of matrix exponential has been found.

We begin by diagonalizing the Hamiltonian [24]:

$$\hat{H}=V\hat{D}V^\dagger \quad \Rightarrow \quad \exp(-i\hat{H}t)=V\exp(-i\hat{D}t)V^\dagger, \tag{29}$$

where $V$ is a matrix of eigenvectors and $\hat{D}$ is a diagonal matrix of eigenvalues. The derivative of the matrix exponential is given by

$$\frac{\partial}{\partial \alpha}\exp\left[i\hat{H}t\right]=\frac{\partial V}{\partial \alpha}\exp\left[i\hat{D}t\right]V^\dagger + V\left[i\frac{\partial \hat{D}}{\partial \alpha}t\right]\exp\left[i\hat{D}t\right]V^\dagger + V\exp\left[i\hat{D}t\right]\frac{\partial V^\dagger}{\partial \alpha} \tag{30}$$

Evaluation of the eigenvalues and eigenvector derivatives has been studied extensively [25-33]. The following sections adapt the approach of Andrew and Tan [34] to the quantum mechanical context.

*Non-degenerate spectrum*

By definition, the $i^{\text{th}}$ eigenvalue $\lambda_i(\alpha)$ of the Hamiltonian $\hat{H}(\alpha)$ and its corresponding eigenvector $|\varphi_i(\alpha)\rangle$ satisfy

$$\lambda_i(\alpha) = \langle \varphi_i(\alpha) | \hat{H}(\alpha) | \varphi_i(\alpha) \rangle \tag{31}$$

$$\langle \varphi_i(\alpha) | \varphi_i(\alpha) \rangle = 1 \tag{32}$$

Differentiating both equations with respect to the parameter $\alpha$ using the product rule and cancelling yields the well known Hellmann-Feynman theorem [35] for the eigenvalue derivative

$$\lambda_i'(\alpha) = \langle \varphi_i(\alpha) | \hat{H}'(\alpha) | \varphi_i(\alpha) \rangle. \tag{33}$$

Once all $\lambda_i'(\alpha)$ are known, the derivative of Equation (31) can be rearranged into

$$\left[\hat{H}(\alpha) - \lambda_i(\alpha)\mathbf{I}\right]|\varphi_i'(\alpha)\rangle = -\left[\hat{H}'(\alpha) - \lambda_i'(\alpha)\mathbf{I}\right]|\varphi_i(\alpha)\rangle \tag{34}$$

in which $\mathbf{I}$ is the identity operator and all terms except $|\varphi_i'(\alpha)\rangle$ are known. The difficulty lies in the fact that the matrix $\hat{H}(\alpha) - \lambda_i(\alpha)\mathbf{I}$ is singular, so the solutions are not unique. This reflects the fact that the phase of eigenvectors is, in general, arbitrary. For example, let $P$ be a diagonal matrix of phase factors $e^{i\phi}$ with unit magnitude. Then if $V = UP$

$$\hat{H} = V\hat{D}V^\dagger = UP\hat{D}P^*U^\dagger = U\hat{D}U^\dagger \tag{35}$$

so both $U$ and $V$ are equally acceptable as eigenvectors. When Equation (29) is solved numerically, $P$ is chosen essentially at random. In order to obtain a well-defined solution to Equation (34), it is simplest to impose an additional constraint [34] on the eigenvectors

$$\langle \varphi_i(\alpha_0) | \varphi_i(\alpha) \rangle = 1 \ \Rightarrow\ \langle \varphi_i(\alpha_0) | \varphi_i'(\alpha) \rangle = 0 \tag{36}$$

for $\alpha$ in a small region around the point $\alpha_0$ where Equation (34) is to be solved. Thus, we solve instead

$$\begin{pmatrix} \hat{H} - \lambda_i \mathbf{I} & -|\varphi_i\rangle \\ \langle \varphi_i | & 0 \end{pmatrix} \begin{pmatrix} |\varphi_i'\rangle \\ \lambda_i' \end{pmatrix} = \begin{pmatrix} -\hat{H}'|\varphi_i\rangle \\ 0 \end{pmatrix} \tag{37}$$

which is well behaved and provides the required eigenvector derivatives.

*Degenerate spectrum*

When there are degenerate eigenvalues at $\alpha_0$, one must consider each block of degenerate eigenvalues together

$$\hat{H}(\alpha)\{|\varphi_i(\alpha)\rangle\} = \{|\varphi_i(\alpha)\rangle\}\Lambda_i(\alpha) \tag{38}$$

where $\{|\varphi_i(\alpha)\rangle\}$ is an $n \times r$ matrix whose columns are the eigenvectors in this $r$-fold degenerate set and $\Lambda_i(\alpha) = \lambda_i(\alpha)\mathbf{I}$. Numerical routines for diagonalizing $\hat{H}$ will return an arbitrary linear combination of these degenerate eigenvectors: $(\{|\varphi_i(\alpha)\rangle\} P^\dagger)$, where $P$ is an as-yet-unknown $r \times r$ unitary matrix. This linear combination may mix eigenvectors corresponding to different eigenvalues derivatives, which would make $\hat{D}'$ not a diagonal matrix. An appropriate choice of $P$ is necessary to prevent this complication.

Now, by the same logic as above

$$\Lambda'_i(\alpha) = P^\dagger \left( P \{\langle \varphi_i(\alpha)|\} \hat{H}'(\alpha_0) \{|\varphi_i(\alpha)\rangle\} P^\dagger \right) P \tag{39}$$

where the $r \times n$ matrix $\{\langle \varphi_i(\alpha)|\}$ is the conjugate transpose of $\{|\varphi_i(\alpha)\rangle\}$ and $\Lambda'_i(\alpha)$ is the diagonal matrix of eigenvalues derivatives. Equation (39) is a small ($r \times r$) eigenproblem, which is easily solved for $\Lambda'_i(\alpha)$ and the true eigenvectors $\{|\varphi_i(\alpha)\rangle\}$ after having computed $\left( P\{\langle \varphi_i(\alpha)|\} \hat{H}'(\alpha_0) \{|\varphi_i(\alpha)\rangle\} P^\dagger \right)$.

The equation analogous to Equation (34) is therefore

$$\left[ \hat{H}(\alpha) - \lambda_i(\alpha)\mathbf{I} \right] \{|\varphi'_i(\alpha)\rangle\} = \{|\varphi_i(\alpha)\rangle\} \Lambda'_i(\alpha) - \hat{H}'(\alpha) \{|\varphi_i(\alpha)\rangle\} \tag{40}$$

To stabilise this equation, we once again use the normalisation condition

$$\{\langle \varphi_i(\alpha_0)|\} \{|\varphi_i(\alpha)\rangle\} = \mathbf{I} \Rightarrow \{\langle \varphi_i(\alpha_0)|\} \{|\varphi'_i(\alpha)\rangle\} = \mathbf{0} \tag{41}$$

and form the block matrix

$$\begin{pmatrix} \hat{H} - \lambda_i \mathbf{I} & -\{|\varphi_i(\alpha)\rangle\} \\ \{\langle \varphi_i(\alpha)|\} & \mathbf{0} \end{pmatrix} \begin{pmatrix} \{|\varphi'_i(\alpha)\rangle\} \\ \Lambda'_i \end{pmatrix} = \begin{pmatrix} -\hat{H}'\{|\varphi_i(\alpha)\rangle\} \\ \mathbf{0} \end{pmatrix} \tag{42}$$

which is easily solved for $\{|\varphi'_i(\alpha)\rangle\}$. The value of $\Lambda'_i$ obtained should be identical to that from diagonalizing Equation (39), which provides a check of the numerical stability of the method at run time.

This algorithm applies equally well to partial derivatives, and can be extended to higher derivatives, beginning by differentiating Equations (33) or (38) several times and solving eigenproblems analogous to Equation (39) for each degenerate block at each stage. Similar procedures are described in [28-29,34].

**Illustrative results**

Figure 1 shows the derivative of the theoretical $^1$H NMR spectrum of strychnine with respect to one of the scalar couplings, calculated using Equation (5). The procedure used amounts to what may be loosely called *co-propagation*: the derivative density matrix is propagated forward in time alongside the main simulation and uses the $\hat{\rho}(t)$ values that the main simulation generates. Figure 2 gives a more sophisticated example (2D COSY), making use of the fact that the *J*-coupling derivatives of a hard pulse are zero and therefore only the time evolution steps need to be differentiated. The unpredictable behaviour of the error resulting from using a finite-difference approximation is illustrated in Figure 1C: as the finite-difference step $\Delta$ gets smaller, the error initially follows the expected $O(h^4)$ scaling. However, numerical noise starts creeping in around $h = 0.01$ Hz and thereafter the accuracy deteriorates rapidly for smaller steps.

As expected, these calculations take much longer if Equation (6) is used to find the parameter derivative, because matrix-matrix multiplications in Liouville space are required. However, if multiple simulations must be carried out from different initial conditions (e.g. the direct dimension of a 2D experiment), this superoperator formulation could be faster, since it would avoid repeated computation of the differentiation superoperators.

The Fourier domain differentiation is significantly faster if only a few points are required in the resulting spectrum (a frequent situation in *e.g.* chemical kinetics fitting where a spectrum needs to be differentiated with respect to the kinetic rate constant). It also has the benefit of not requiring apodization. An example of such a derivative is given in Figure 3.

A less obvious differentiation parameter – waveform truncation level – is demonstrated in Figure 4. The resulting derivative gives a measure of stability of the magnetic resonance experiment with respect to the clipping of the wings of a Gaussian pulse. As intuitively expected, waveform clipping has no effect if the signal is positioned exactly on resonance, but does introduce imperfections into the excitation of off-resonance peaks.

Figure 5 demonstrates eigensystem differentiation using a simple model Hamiltonian

$$\hat{H}(\alpha) = \begin{pmatrix} \cos(\alpha) & \sin(\alpha) & 0 \\ -\sin(\alpha) & \cos(\alpha) & 0 \\ 0 & 0 & 1 \end{pmatrix} \begin{pmatrix} 4-\alpha^2 & 0 & 0 \\ 0 & 10 & 0 \\ 0 & 0 & 3\alpha \end{pmatrix} \begin{pmatrix} \cos(\alpha) & -\sin(\alpha) & 0 \\ \sin(\alpha) & \cos(\alpha) & 0 \\ 0 & 0 & 1 \end{pmatrix} \quad (43)$$

whose eigenvectors and eigenvalues are known analytically. Notice that there are no difficulties for degenerate eigenvalues at $\alpha = 10/3$, 1 or $-4$. Most importantly, all elements of the matrix exponential $e^{i\hat{H}}$ are accurate to within an absolute error of $10^{-13}$ throughout this range. Similar tests with $3\times 3$ and $4\times 4$ complex model Hamiltonian produced comparable accuracy in the all important matrix exponentials.

Figure 6 is a more sophisticated example from low-field electron paramagnetic resonance of radical pairs [25,36]. In such systems, the reaction yield is governed by the strength of an applied magnetic field. The simplest radical pair comprises two electrons, and a single ¹H nucleus in one of the radicals and has Hamiltonian

$$\hat{H}(B_0) = -\gamma_e B_0 \left(\hat{S}_{Az} + \hat{S}_{Bz}\right) + a\hat{S}_A \cdot \hat{I}. \quad (44)$$

Neglecting the reaction kinetics for now, the probability of a radical pair created in a pure electronic singlet state remaining in that state is

$$S(t; B_0) = \text{Tr}\left[e^{-i\hat{H}t} \hat{\rho}_0 e^{i\hat{H}t} \hat{P}_S\right] = \text{Tr}\left[e^{-i\hat{D}t}\left(\hat{V}^\dagger \hat{\rho}_0 \hat{V}\right)e^{i\hat{D}t}\left(\hat{V}^\dagger \hat{P}_S \hat{V}\right)\right] \quad (45)$$

where $\hat{D}$ is a diagonal matrix containing Hamiltonian eigenvalues, $\hat{V}$ is the corresponding eigenvector matrix, $\hat{P}_S$ is the electron singlet projection operator and the initial density matrix $\hat{\rho}_0 = \hat{P}_S/2$ because the initial nuclear spin state is random. Using the "exponential model" [36-39] for reaction kinetics, the total yield of singlet product

$$\Phi_S(B_0) = \int_0^\infty S(t; B_0) k e^{-kt} dt. \quad (46)$$

Here, Equation (46) may be integrated analytically if $\hat{H}$ has been diagonalised, which more than compensates for the effort required for diagonalisation. In this case, the method of eigensystem differentiation is therefore the most attractive.

Figure 6 compares $d\Phi_S(B_0)/dB_0$ computed by eigensystem differentiation with an exact calculation made analytically for specific values of the hyperfine constant $a$ and rate constant $k$. Agreement is essentially exact throughout the range that was plotted.

## Conclusion

In simulations of spin dynamics, the formalism outlined above provides a straightforward, accurate and numerically stable avenue to determine derivatives with respect to spin system or pulse sequence parameters. The resulting derivatives may be used in fitting, optimization, performance evaluation and stability analysis of spin dynamics experiments and experimental results.

## Acknowledgements

This work was supported by EPSRC (EP/F065205/1, EP/H003789/1) and NIS (PHY05-51164). The authors are grateful to Niels Chr. Nielsen and Alan Andrew for helpful advice. IK gratefully acknowledges the Kavli Institute for Theoretical Physics at UCSB. CTR is funded by NIHR Biomedical Research Centre Programme and Merton College, Oxford.

# References

[1] A. D. Bain, D. M. Rex, and R. N. Smith, Magn. Reson. Chem. **39** (3), 122 (2001).

[2] L. C. Wang, A. V. Kurochkin, and E. R. P. Zuiderweg, J. Magn. Reson. **144** (1), 175 (2000).

[3] S. Dusold, E. Klaus, A. Sebald, M. Bak, and N. C. Nielsen, J. Am. Chem. Soc. **119** (30), 7121 (1997).

[4] H. J. M. Degroot, S. O. Smith, A. C. Kolbert, J. M. L. Courtin, C. Winkel, J. Lugtenburg, J. Herzfeld, and R. G. Griffin, J. Magn. Reson. **91** (1), 30 (1991).

[5] S. Castellano and A. A. Bothner-By, J. Chem. Phys. **41** (12), 3863 (1964).

[6] Z. Tosner, T. Vosegaard, C. Kehlet, N. Khaneja, S. J. Glaser, and N. C. Nielsen, J. Magn. Reson. **197** (2), 120 (2009).

[7] I. I. Maximov, Z. Tosner, and N. C. Nielsen, J. Chem. Phys. **128** (18) (2008).

[8] N. Khaneja, T. Reiss, C. Kehlet, T. Schulte-Herbruggen, and S. J. Glaser, J. Magn. Reson. **172** (2), 296 (2005).

[9] T. E. Skinner, T. O. Reiss, B. Luy, N. Khaneja, and S. J. Glaser, J. Magn. Reson. **163** (1), 8 (2003).

[10] J. Nocedal and S. J. Wright, *Numerical optimization*, 2nd ed. (Springer, New York ; London, 2006).

[11] N. I. M. Gould, D. Orban, and P. L. Toint, *Numerical methods for large-scale nonlinear optimization*. (Council for the Central Laboratory of the Research Councils, Didcot, 2004).

[12] R. Dunkel, C. L. Mayne, J. Curtis, R. J. Pugmire, and D. M. Grant, J. Magn. Reson. **90** (2), 290 (1990).

[13] J. A. Nelder and R. Mead, Comp. J. **7** (4), 308 (1965).

[14] D. A. Pearlman, J. Biomol. NMR **8** (1), 49 (1996).

[15] D. R. Yarkony, *Modern electronic structure theory*. (World Scientific, London, 1995).

[16] P. Pulay, Mol. Phys. **17** (2), 197 (1969).

[17] A. F. Bartelt, M. Roth, M. Mehendale, and H. Rabitz, Phys. Rev. A **71** (6) (2005).

[18] F. Shuang and H. Rabitz, J. Chem. Phys. **121** (19), 9270 (2004).

[19] R. R. Ernst, G. Bodenhausen, and A. Wokaun, *Principles of nuclear magnetic resonance in one and two dimensions*. (Clarendon, Oxford, 1987).

[20] T. C. Fung, Int. J. Numer. Meth. Eng. **59** (10), 1273 (2004).

[21] C. Moler and C. Van Loan, SIAM Rev. **45** (1), 3 (2003).

[22] I. Najfeld and T. F. Havel, Adv. Appl. Math. **16** (3), 321 (1995).

[23] G. H. Golub and C. F. Van Loan, *Matrix computations*, 2nd ed. (The Johns Hopkins University Press, Baltimore, 1989).

[24] J. H. Wilkinson, *The algebraic eigenvalue problem*. (Clarendon Press, Oxford,, 1965).

[25] U. E. Steiner and T. Ulrich, Chem. Rev. **89** (1), 51 (1989).

[26] A. L. Andrew, K. W. E. Chu, and P. Lancaster, SIAM J. Matrix Anal. Appl. **14** (4), 903 (1993).

[27] A. L. Andrew, Comp. Tech. Appl., 51 (1998).

[28] A. L. Andrew and R. C. E. Tan, Comm. Num. Meth. Eng. **15** (9), 641 (1999).

[29] A. L. Andrew and R. C. E. Tan, Num. Lin. Alg. Appl. **7** (4), 151 (2000).

[30] A. L. Andrew and R. C. E. Tan, in *Linear operators and matrices : the Peter Lancaster anniversary volume*, edited by I. Gohberg and H. Langer (Birkhauser Verlag, Boston, 2002), Vol. 130, pp. 43.

[31] K.-W. E. Chu, SIAM J. Num. Anal. **27** (5), 1368 (1990).

[32] Z. H. Xu and B. S. Wu, Int. J. Num. Meth. Eng. **75** (8), 945 (2008).

[33] P. Lancaster, Numerische Mathematik **6**, 377 (1964).

[34] A. L. Andrew and R. C. E. Tan, SIAM J. Matrix Anal. Appl. **20** (1), 78 (1998).

[35] R. P. Feynman, Phys. Rev. **56** (4), 340 (1939).

[36] C. T. Rodgers, Pure Appl. Chem. **81** (1), 19 (2009).

[37] R. Kaptein and J. L. Oosterhoff, Chem. Phys. Lett. **4** (4), 195 (1969).

[38] R. G. Lawler and G. T. Evans, Industrie Chimique Belge **36** (12), 1087 (1971).

[39] B. Brocklehurst and K. A. McLauchlan, Int. J. Rad. Biol. **69** (1), 3 (1996).

[40] I. Kuprov, J. Magn. Reson. **195** (1), 45 (2008).

[41] I. Kuprov, N. Wagner-Rundell, and P. J. Hore, J. Magn. Reson. **189** (2), 241 (2007).

**Figure captions**

**Figure 1.** *A:* Fragments of the theoretical 90°–acquire $^1$H NMR spectrum of strychnine. Simulations were performed for a 22-spin system (with magnetic parameters obtained from a separate GIAO B3LYP/cc-pVDZ calculation using Gaussian03), employing state space restriction [40-41] to *k*-spin orders around every spin, where *k* is the number of *J*-coupled neighbors (or *k*=4 if that is larger), and with a diagonal DD-CSA relaxation superoperator. *B:* Analytical derivative of the same spectrum with respect to one of the scalar couplings (solid line) and numerical derivative with respect to the same parameter (circles) using an $O(h^4)$ four-point central finite difference approximation. Wall clock times for propagation with analytical derivatives: 7.6 s; and for the numerical derivatives: 29.5 s. *C:* Maximum absolute difference between the analytical derivative and the finite-difference approximation as a function of the finite-difference step size.

**Figure 2.** *A:* Fragment of the $^1$H–$^1$H COSY spectrum of strychnine. Simulations were performed as described in Figure 1. *B:* Analytical derivative of the same COSY spectrum with respect to one of the scalar couplings. Wall clock times for propagation: original spectrum: 12 min; analytical derivative: 15 min. For comparison the equivalent $O(h^4)$ four-point central finite difference approximation took 49 min.

**Figure 3.** *A:* $^1$H NMR spectra of a strongly coupled two-spin system undergoing symmetric chemical exchange with rate constants varying between $k \ll J, \Delta\omega$ (bottom trace) and $k \gg J, \Delta\omega$ (top trace). *B:* Analytical derivatives of the same spectra with respect to the exchange rate constant *k*. (Each spectrum was scaled vertically to make its features visible.)

**Figure 4.** Derivative of the excitation profile of a 90° Gaussian pulse with respect to waveform truncation. *A:* The function $f(x) = \tanh(k(\alpha - x)) + \tanh(k(\alpha + x))$ used to approximate the waveform truncation. The three traces correspond to $k = 10, 10^2, 10^3$. *B:* Gaussian waveform showing the truncation parameter (in units of standard deviation). *C:* T*op*: Magnitude-mode 1D NMR spectrum of a linear chain of 33 protons with regularly spaced chemical shifts and equal nearest-neighbor *J*-couplings *C:* Bottom: Analytical derivative of the spectrum above with respect to the truncation parameter $\alpha$ when $\alpha = 3.0$.

**Figure 5.** Comparison of the derivatives of the eigenvalues and eigenvectors of $\hat{H}(\alpha)$ defined in Equation (43). Exact, analytical solutions are shown as lines, numerical solutions from the eigensystem differentiation algorithm are shown as crosses. Notice the good behaviour when eigenvalue degeneracies occur (at $\alpha = -4$, 1 and 10/3). *Left:* First derivative with respect to $\alpha$ of the eigenvalues of $\hat{H}(\alpha)$. *Right:* Absolute values of each component of the eigenvector derivatives with respect to $\alpha$. The absolute value is plotted to mitigate the arbitrary phase factors in the numerical solutions.

**Figure 6.** Derivative of singlet yield with respect to field strength in a radical pair with hyperfine coupling $a$ to a single spin-1/2 nucleus, using the exponential model with rate constant $k = a/10$. Numerical results (crosses) were obtained from Equation (46) using the eigensystem differentiation algorithm, exact values (line) were obtained using a computer algebra system to give Equations (S1) – (S3) in the Supporting Information.

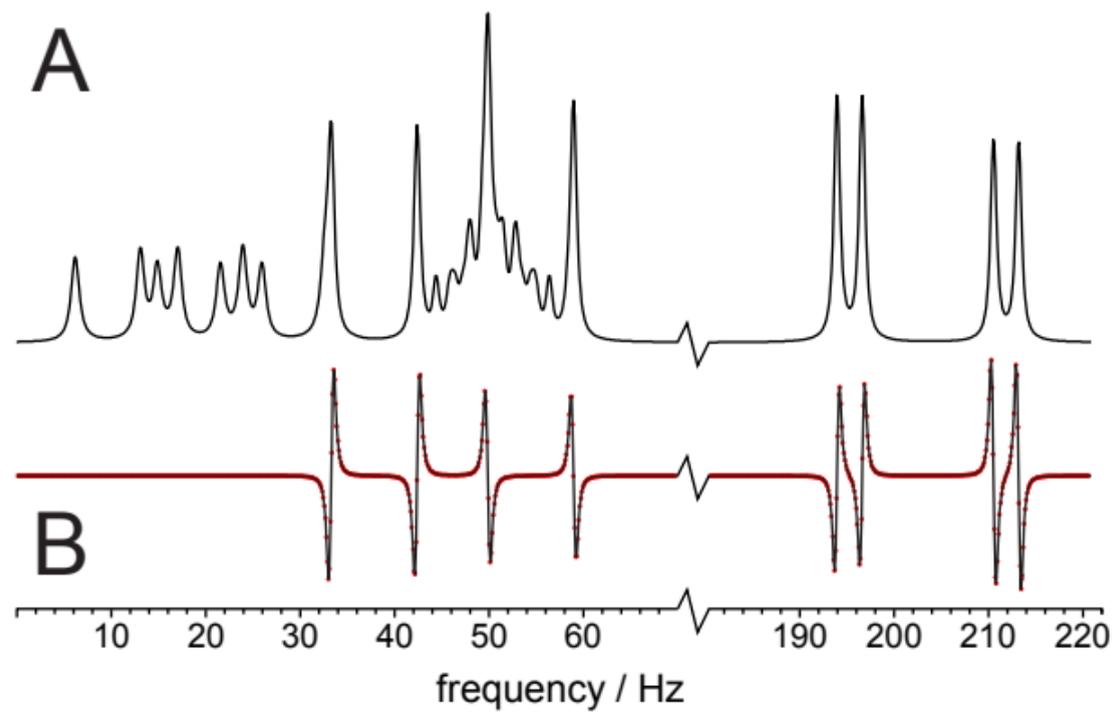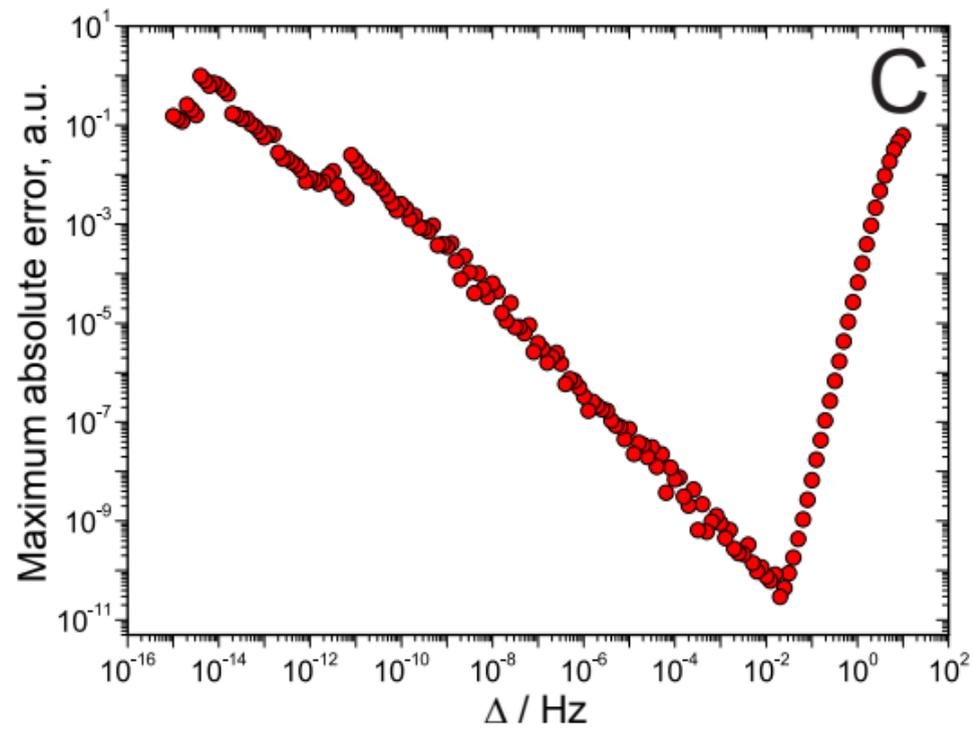

Figure 1

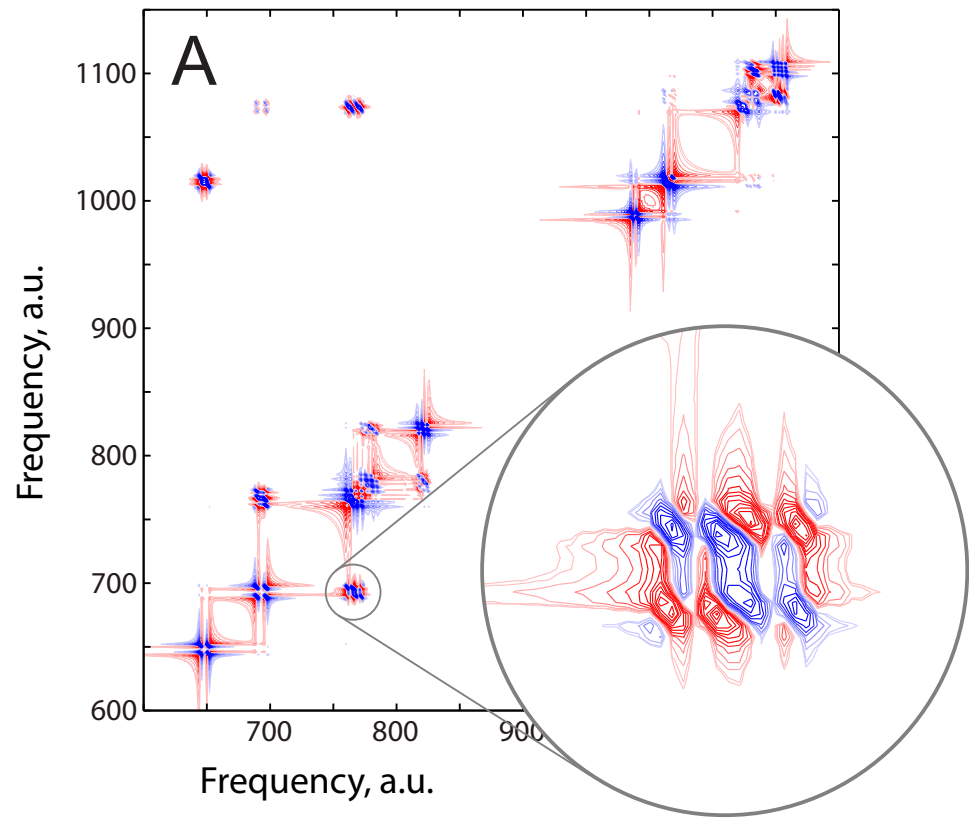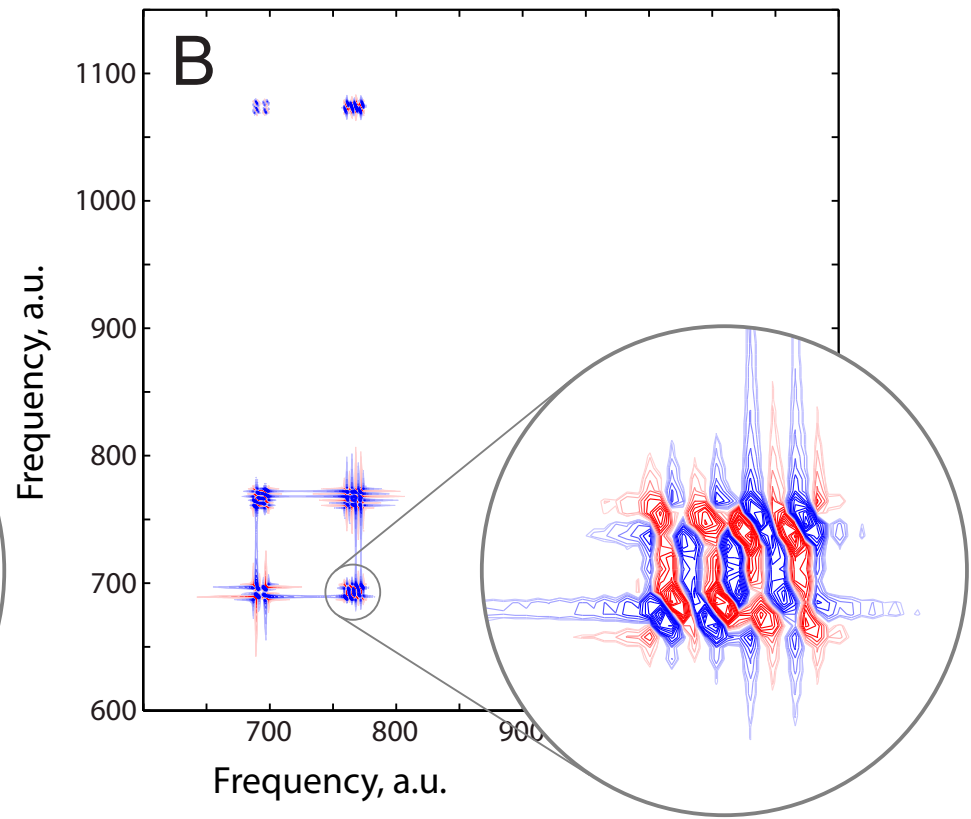

Figure 2

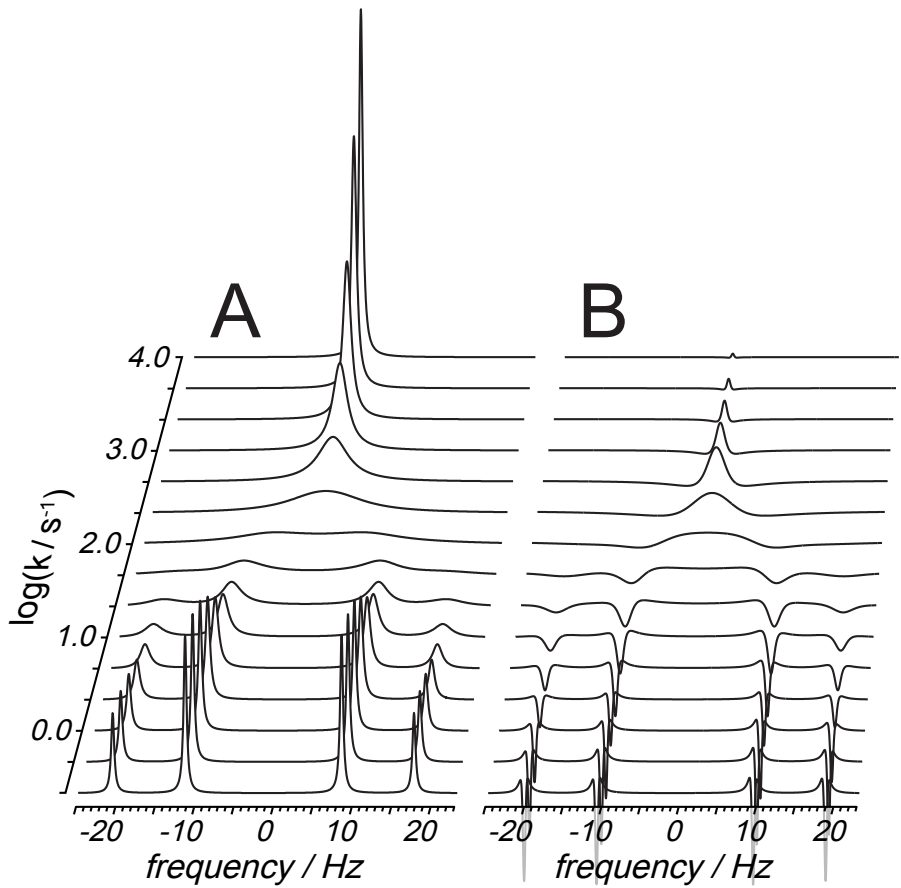

Figure 3

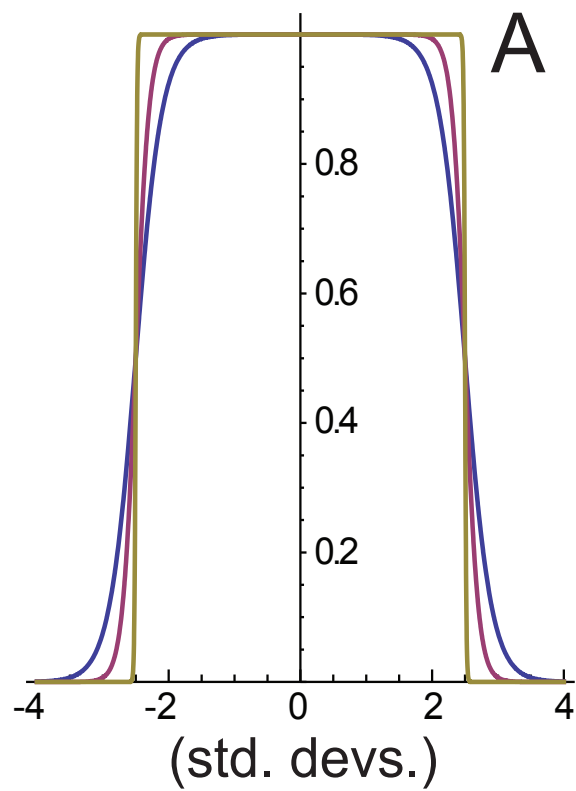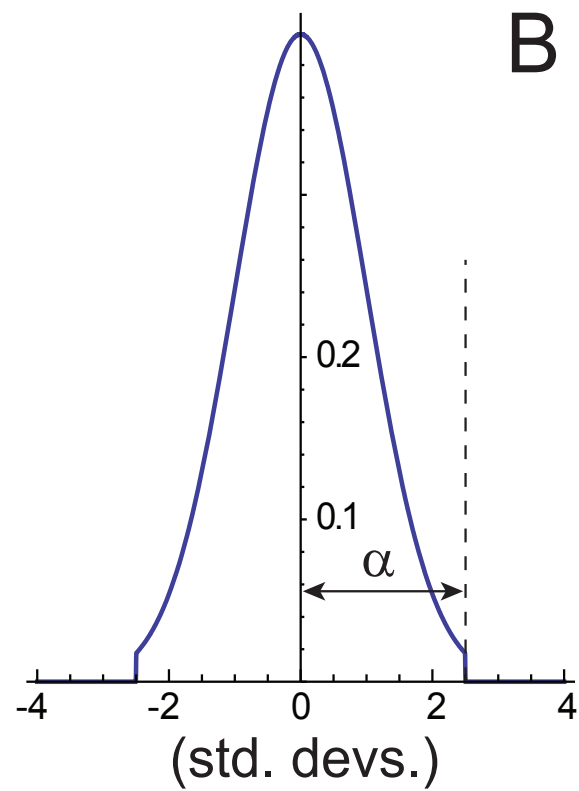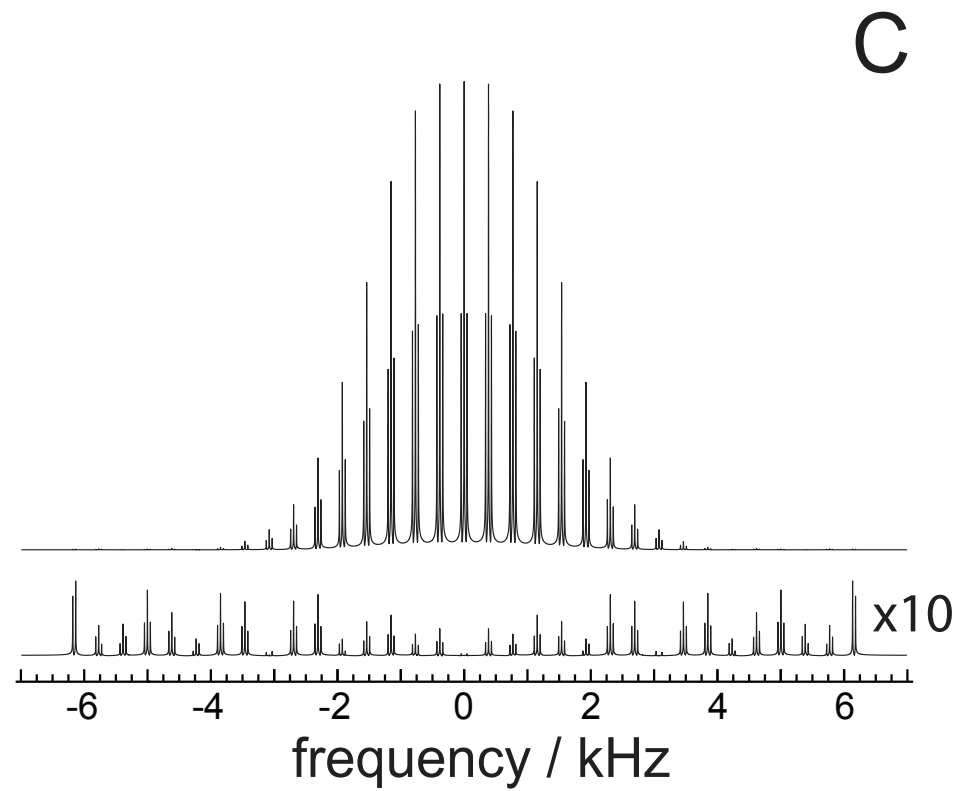

Figure 4

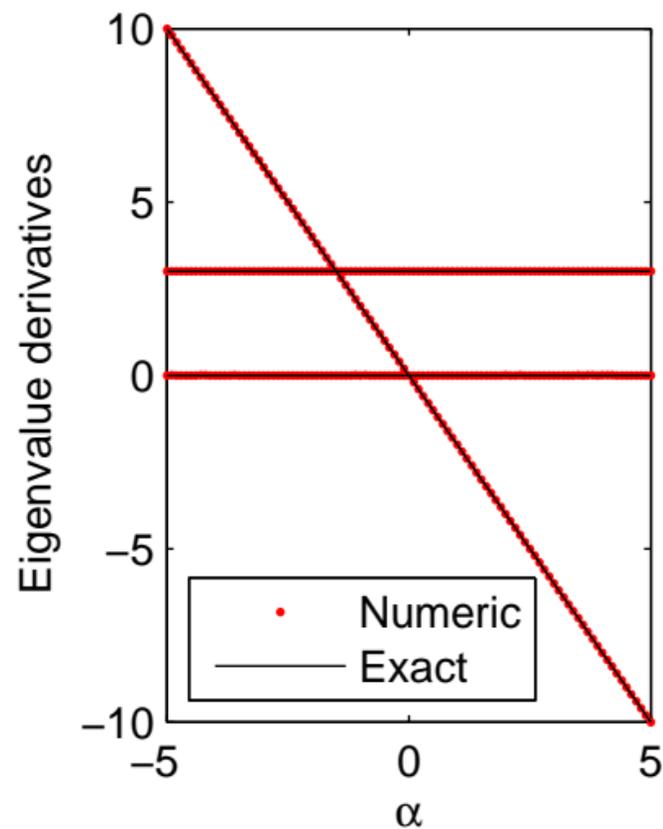
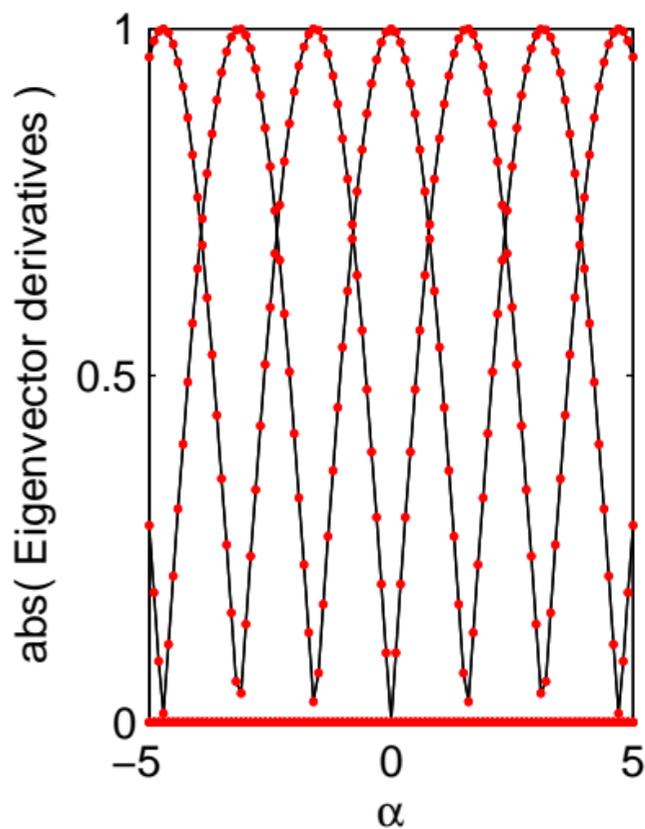

Figure 5

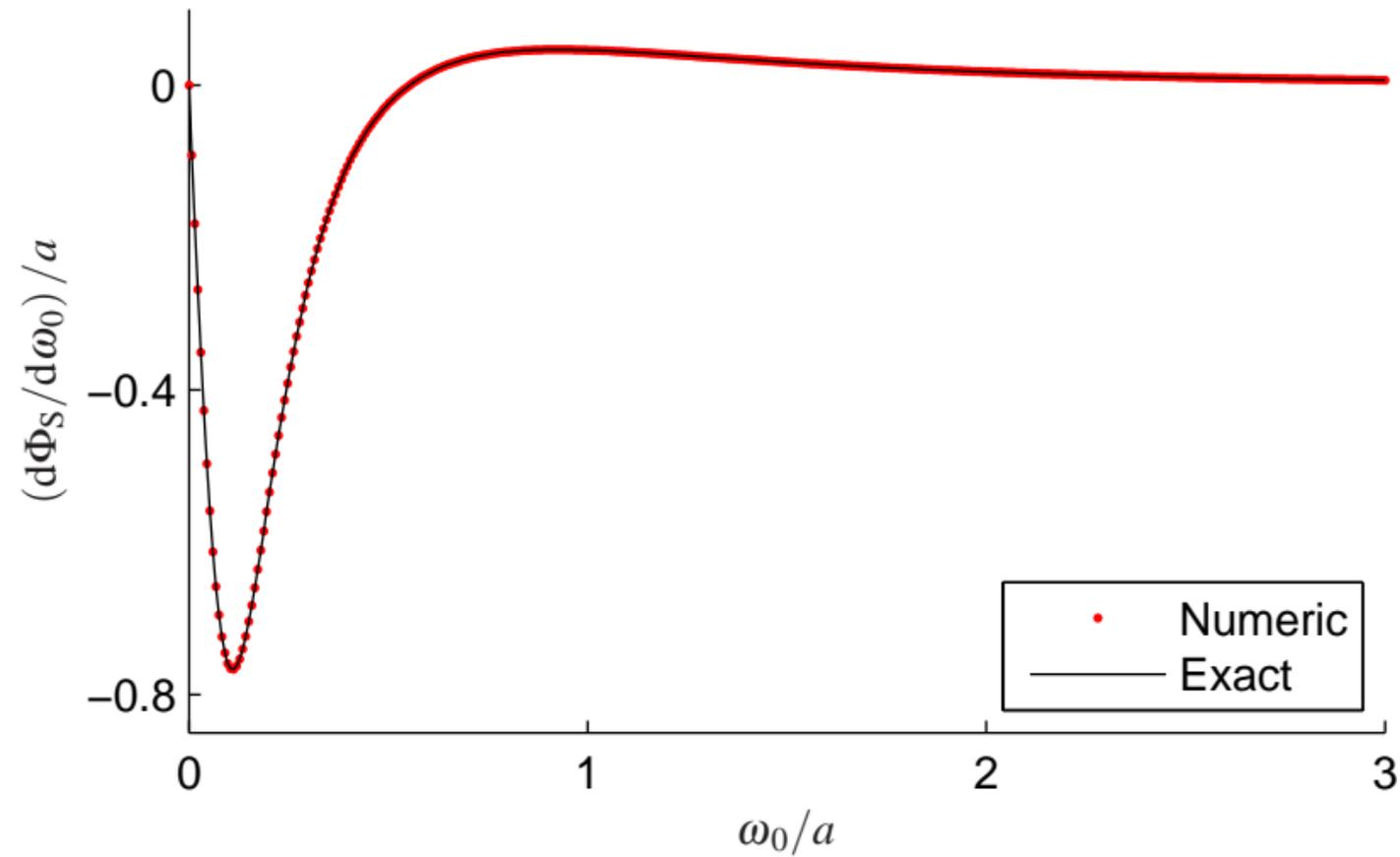

Figure 6